\documentclass{aa}
\usepackage[dvips]{graphicx}
\usepackage{psfig}
\def\HII{H{\sc ii}}

\def\UC{UC~H{\sc ii}}

\def\kms{\mbox{km~s$^{-1}$}}

\def\Log{\mbox{\rm Log}}

\def\ace{CH$_{3}$C$_{2}$H}

\begin{document}
\title{The structure of molecular clumps around
high-mass young stellar objects
 \thanks{Based on observations carried out with the IRAM Pico Veleta telescope.
 IRAM is supported by INSU/CNRS (France), MPG (Germany) and IGN (Spain).}}
\author{F. Fontani \inst{1} \and R. Cesaroni \inst{2}
	\and P. Caselli \inst{2} \and L. Olmi \inst{3}}
\institute{Dipartimento di Astronomia e Fisica dello spazio, Largo E. Fermi 2,
           I-50125 Firenze, Italy \and
	   Osservatorio Astrofisico di Arcetri, Largo E. Fermi 5,
           I-50125 Firenze, Italy
\and
           LMT/GTM Project, Dept. of Astronomy, 815J Lederle GRT Tower B,
 University of Massachusetts, 710 N. Pleasant st., Amherst, MA 01003, USA
 }
\offprints{F. Fontani, \email{fontani@arcetri.astro.it}}
\date{Received 14 December 2001; Accepted 12 April 2002}

\titlerunning{\ace\ in UC \HII\ regions}
\authorrunning{Fontani et al.}

\abstract{
We have used the IRAM 30-m and FCRAO 14-m telescopes to observe the molecular clumps
 associated with 12 ultracompact (UC) \HII\ regions in the $J$=6--5, 8--7 and
 13--12 rotational transitions of methyl-acetylene (\ace). Under the
 assumption of LTE and optically thin emission, we have derived
 temperature estimates ranging from 30 to 56 K. 
We estimate that the clumps have diameters of 0.2--1.6~pc,
H$_2$ densities of $10^5$--$10^6~\rm{cm^{-3}}$, and masses of
$10^2$--$2\,10^4~M_\odot$. We compare these values with those obtained by
other authors from different molecular tracers and find that the
H$_2$ density and the temperature inside the clumps vary respectively like
$n_{\rm H_2}\propto R^{-2.6}$ and $T\propto R^{-0.5}$, with $R$ distance
from the centre.
We also find that the virial masses of the clumps are
$\sim$3 times less than those derived from the \ace\ column densities: we
show that a plausible explanation is that magnetic fields play an important
role to stabilise the clumps, which
are on the verge of gravitational collapse.
Finally, we show that the \ace\ line width increases for decreasing
distance from the clump centre: this effect is consistent with infall
in the inner regions of the clumps. We conclude that the clumps around UC \HII\
regions are likely to be transient ($\sim$$10^5$~yr) entities, remnants of
isothermal spheres currently undergoing gravitational collapse: the high mass
accretion rates ($\sim$$10^{-2}~M_\odot$~yr$^{-1}$) lead to massive star
formation at the centre of such clumps.
\keywords{Stars: formation -- Radio lines: ISM -- ISM: molecules}
}

\maketitle

\section{Introduction}

While substantial progress has been done in recent years towards a better
description
of the formation of low-mass stars, much remains to be understood of their
massive counterparts. In fact, several problems hinder the study of high-mass
star formation: massive stars ($M\geq 10~M_\odot$) are more
distant, interact more strongly with their environment,
and have shorter evolution timescales.
After the protostellar phase, they reach the
ZAMS and create an \HII\ region in the circumstellar medium before the 
accretion phase is finished. When the star is very young (i.e. $\leq10^{5}$~yr),
this ionised
region is small (i.e. $\leq0.1$ pc for an O star) with respect to the typical 
diameter of an \HII\ region associated with a more evolved OB star (i.e. 
$\sim$1--10~pc), and is named ultracompact (UC) \HII\ region. 
 Given the strong interaction between early type stars and their surroundings,
 it is important to have knowledge of the physical 
characteristics of their parental molecular clouds.
With this in mind, Churchwell et al. (1990),
Cesaroni et al. (1991), and Hofner et al. (2000) observed respectively
NH$_3$, C$^{34}$S, and C$^{17}$O towards a few UC \HII\ regions,
thus assessing the existence of molecular clumps surrounding
such UC \HII\ regions. They estimated
masses of $10^3$--$10^4~M_\odot$, densities of $10^4$--$10^5$~cm$^{-3}$,
and diameters of 0.4--1~pc. In another study,
Olmi et al.~(1993) detected various    
CH$_3$CN rotational lines in a limited sample of UC \HII\ regions, which
demonstrated the existence of even hotter, denser cores inside the more
extended clumps observed in the C$^{34}$S lines.
Notwithstanding these results, a robust temperature estimate for the
molecular clumps surrounding UC \HII\ regions was still missing. In fact,
on the one hand tracers such as NH$_3$ and C$^{17}$O provide temperature
estimates based on 2--3 transitions only and hence prone to relatively
large uncertainties; on the other, CH$_3$CN is an excellent ``thermometer''
but arises from a region much smaller and denser than the 1~pc clumps seen
in C$^{34}$S and C$^{17}$O.
Hence, the main goal of the present study was to derive a good temperature  
estimate for the medium density molecular surroundings of UC \HII\ regions.

The \ace\ species is especially suitable to our purposes. This is a
symmetric-top molecule (see Townes \& Schawlow 1975) with a small dipole
moment ($\mu=0.75$ Debye, Dubrulle et al. 1978), which makes thermalisation
easy even at densities as low as 10$^4$~cm$^{-3}$. We thus expect \ace\ to be
in ``local thermodynamical equilibrium'' (LTE) conditions in the clumps found
by Cesaroni et al. (1991)
around the UC \HII\ regions. Moreover, the large number of transitions
observable in the same bandwidth makes the temperature estimate very
accurate.
For instance, Bergin et al. (1994) successfully used this method to measure the
temperature of dense cores in giant molecular clouds, deriving
temperatures of $\sim 40$~K.

In Sect.~\ref{sobs} we describe the observations and data reduction procedure,
in Sect.~\ref{sres} we summarise the observational results and
derive the physical parameters of the clumps, in Sect.~\ref{sdisc}
we discuss the implications of the results obtained.
The conclusions are drawn in Sect.~\ref{sconc}.

\section{Observations and data reduction}
\label{sobs}

\subsection{IRAM 30-m telescope}
\label{siram}

The data from the IRAM 30-m telescope 
were obtained in the period from 
August 24 to 29, 1997. The source list is shown in Table~1. The velocity
of each source listed in Table 1 is taken from previous observations of 
various molecular transitions, and it is only an indicative value of the real
velocity of the gas from which the \ace\ lines arise. We 
simultaneously observed the (6--5), (8--7) and (13--12) rotational transitions
of \ace\ in the 3~mm, 2~mm and 1.3~mm bands respectively.
The half-power beam width (HPBW)
of the IRAM 30-m telescope was 22\arcsec, 17\arcsec, and 
12\arcsec\ at 3~mm, 2~mm and 1.3~mm, respectively.

The alignment between the different receivers was checked through continuum
cross scans on Jupiter at the three frequencies of observation, and found to be
accurate to within 4\arcsec. Pointing was checked every 0.5--1~hours by means
of continuum cross scans on some \UC\ regions of the Churchwell et al. (1990)
sample. We believe it is accurate to within~4\arcsec. 

\begin{table}
\begin{center}
\caption{List of the observed sources}
\label{tsou}
\begin{tabular}{rlccc}
\hline 
\# & Source   &   $\alpha$(B1950)  &  $\delta$(B1950)  &  $v_{\rm LSR}$ \\
   &          &   (h m s) & (\degr\ \arcmin\ \arcsec) & (\kms) \\
\hline 
 1 & G5.89$-$0.39 &   17 57 26.8 &  $-$24 03 56  & 10.0      \\
 2 & G9.62+0.19  & 18 03 16.2  & $-$20 32 03   & 4.4       \\
 3 & G10.47+0.03 &  18 05 40.3 & $-$19 52 21   & 67.8      \\
 4 & G10.30$-$0.15 $(^*)$ &  18 05 57.7  & $-$20 06 26   & 13.5    \\
 5 & G10.62$-$0.38 &  18 07 30.7 & $-$19 56 29   & $-$3.1      \\
 6 & G19.61$-$0.23 &  18 24 50.3 & $-$11 58 33   & 41.6      \\
 7 & G29.96$-$0.02 &  18 43 27.1 & $-$02 42 36   & 98.0      \\
 8 & G31.41+0.31 &  18 44 59.2 & $-$01 16 07   & 97.0      \\
 9 & G34.26+0.15 &  18 50 46.1 & +01 11 12    & 58.0      \\
10 & G45.47+0.05  & 19 12 04.3 & +11 04 11    & 62.0     \\
11 & W51D $(^*)$    &  19 21 22.3 & +14 25 15    & 60.0     \\
12 & IRAS\,20126+4104 &   20 12 41.0 &  +41 04 21   & $-$3.5      \\
\hline
\end{tabular}
\end{center}
 $(^*)$ not observed with the FCRAO telescope
\end{table}

\begin{table}
\begin{center}
\caption{Spectrometers used for the IRAM 30-m observations}
\begin{tabular}{c|c|c|c}
\hline 
  Band   &  Spectrometer   & Channel Spacing & Total Band \\
          &                 &    (MHz)  & (MHz) \\
\hline
 3~mm    &  filterbank    &   1.0  & 256 \\
         &  autocorrelator   &  0.078 &  70 \\ \hline
 2~mm    & filterbank      & 1.0  & 256 \\
       &  autocorrelator    & 0.078 &  70 \\ \hline
 1.3~mm   &  filterbank    & 1.0  & 512 \\
       &  autocorrelator   & 0.078  & 140 \\ \hline
\end{tabular}
\end{center}
\end{table}    
  
The spectrometers used simultaneously are an autocorrelator,
which covered only the lowest $K$ components with high spectral resolution, and
a filterbank with low spectral resolution, covering all $K$ lines for the
(6--5) and (8--7) transitions and up to the $K$=10 component for the (13--12)
transition.
The main parameters of the spectrometers are listed in Table~2.

For all sources, the observations were made towards the nominal central
positions given in Table~1. Only towards G31.41, we obtained two 3$\times$3
point maps, one with 12\arcsec\ and the other with 24\arcsec\ spacing.

We observed using ``wobbler switching'', namely a nutating secondary reflector
with a beam-throw of
240\arcsec\ in azimuth and a phase duration of 2~sec.
The data were calibrated with the chopper wheel technique (see Kutner \&
Ulich 1981).
We checked the calibration by measuring the intensities of two planets
observed at the same frequencies as our sources. 
The main beam brightness temperature, $T_{\rm MB}$, and the flux density,
$F_\nu$, are related by the expression
$F_\nu({\rm Jy})=4.7~T_{\rm MB}({\rm K})$.


\subsection{FCRAO Telescope}

Observations of \ace\ (6--5) were carried out between March 16 and 
May 18 1997 
with the Five College Radio Astronomy Observatory (FCRAO), located in New Salem 
(Massachusetts, USA). The receiver used was the
QUARRY focal plane array (Erickson et al. 1992), which covered the  
3mm band (90--116~GHz) and consisted of
15 separate beams disposed in a 3$\times$5 configuration.  
The telescope HPBW at the frequency of the \ace(6--5) line
is 50\arcsec. The calibration was carried out using the chopper wheel technique.
The observations were made in position switching,
moving the antenna in azimuth by 2\arcmin.
The main beam brightness temperature, $T_{\rm MB}$, and the flux density,
$F_\nu$, are related by the expression
$F_\nu({\rm Jy})=21.5~T_{\rm MB}({\rm K})$.

The spectrometer used was an autocorrelator, with 512 channels and a spectral
resolution of 80~kHz.
At FCRAO we observed the same sample of UC \HII\ regions
observed with the 30-m telescope, except G10.30 and W51D.

\begin{figure*}
\centerline{\psfig{figure=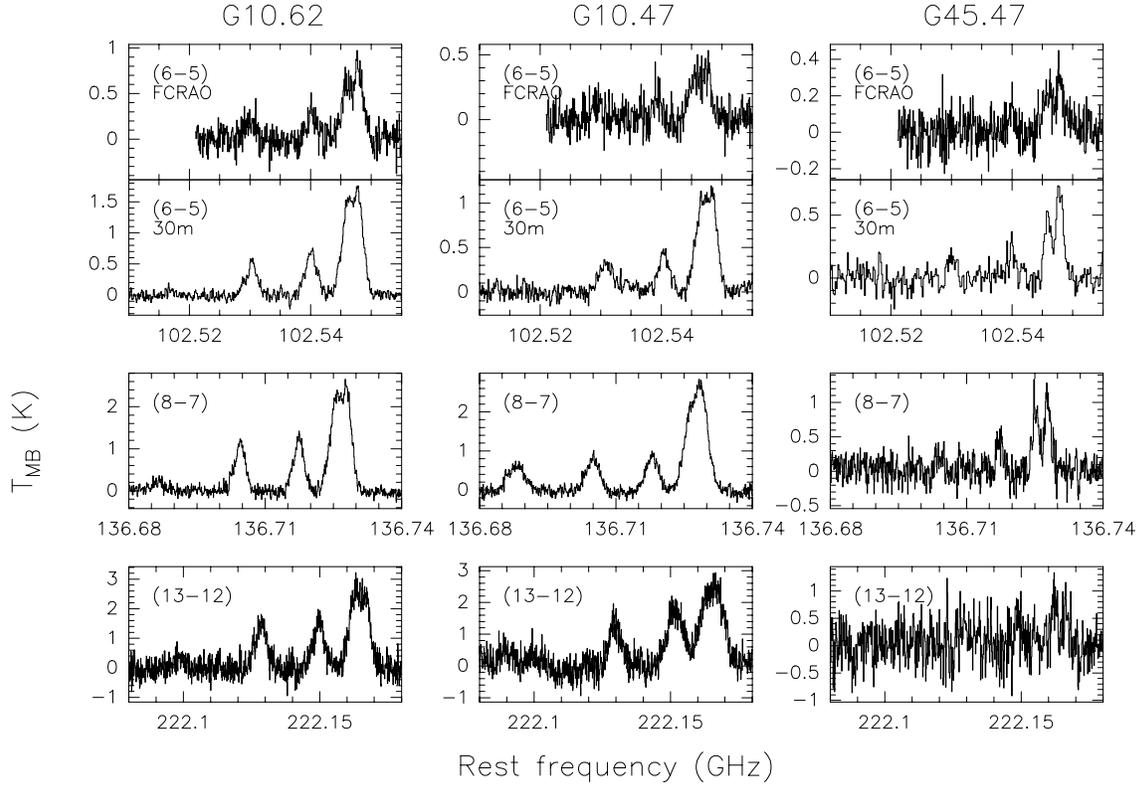,angle=-90,width=15cm}}
\caption{Spectra of the \ace\ (6--5), (8--7), and (12--11) rotational
 transitions towards sources G10.62, G10.47 and
G45.47. On top, the FCRAO (6--5) spectra are shown}
\end{figure*}

\subsection{Data reduction and fitting procedure}
\label{sfit}

\ace\ is a symmetric-top molecule, with
rotational levels described by two quantum numbers: $J$ is associated with 
the total angular momentum, whereas $K$ is associated with its projection on the 
symmetry axis. For each value of $J$, the selection rules $\Delta J=\pm1$
and $\Delta K=0$ make possible only rotational radiative transitions 
$J+1\rightarrow J$
with $K\leq J$; consequently, for each $J+1\rightarrow J$ transition,
only $J+1$ $K$-lines can be seen. In our observations
we could detect only lines up to a maximum of $K=4$.

The data reduction made use of the CLASS program of the GAG software 
developed at IRAM and Observatoire de Grenoble. 
In order to derive the line
parameters of \ace\ we assumed that all the $K$ components of each
$J+1\rightarrow J$ transition arise from the same region, and are hence
characterised by the same peak velocity and line width. Then, Gaussian  
fits to each $J+1\rightarrow J$ spectrum were made with line separations
between different $K$ components fixed to the laboratory values and line
widths assumed to be identical.

\section{Results}
\label{sres}

With the IRAM 30-m telescope we detected \ace\ (6--5), (8--7) and (13--12)
towards
all of the 12 sources of our sample, whereas with the FCRAO we detected \ace\
(6--5) towards all of the sources but G19.61.
 
In Tables 3, 4, 5 and 6 we list the line parameters obtained with the data
reduction procedure described in Sect.~\ref{sfit}.
In all cases, the difference between the line profile and the
Gaussian fit was within the noise.
For each source we list: the LSR velocity, $V_{\rm LSR}$;
the line full width at half maximum (FWHM), $\Delta v_{1/2}$;
the emission integrated under each line, $\int T_{\rm MB}{\rm d}v$;
and the 1$\sigma$ rms noise
of the spectrum. For Gaussian profiles, the peak temperatures
$T_{\rm MB}^{\rm max}$
are related to $\int T_{\rm MB}{\rm d}v$ by the relationship:
\begin{equation}
\int T_{\rm MB}{\rm d}v=\frac{\Delta v_{1/2}}{2 \sqrt{\frac{\ln2}{\pi}}}T_{\rm MB}^{\rm max}\;
\end{equation}
The errors quoted do not take into account the calibration uncertainties, but
represent only the formal errors of the fit procedure.
The calibration errors will be discussed in Sect.~\ref{stempden}.

\begin{table*} 
\begin{center}
\caption{Parameters of the \ace(6--5) lines observed with the IRAM 30-m
telescope. The results given here are
based on the autocorrelator (high resolution) spectra. The errors are
given by the fit procedure, and do not consider the calibration error
on $T_{\rm MB}$.
$\Delta v_{1/2}$ represents the full width at half maximum (FWHM) of the line.
$\sigma$ is the noise level in K}
\begin{tabular}{l|cccccccc}
\hline
Source  & $v_{\rm LSR}$ &  $\Delta v_{1/2}$ &  \multicolumn{5}{c}{$\int T_{\rm MB}{\rm d}v$ (K \kms)}  &   $\sigma$ \\
\cline{4-8}
          & (\kms)  &  (\kms)     & $K=0$ & 1 & 2 & 3 & 4 &  (K)    \\
\hline
G5.89     & 8.54$\pm$0.08 & 3.45$\pm$0.1  &  15.15$\pm$0.1 & 14.07$\pm$0.08 & 
  7.24$\pm$0.1  & 5.35$\pm$0.1 & 0.85$\pm$0.1  & 0.1    \\
G9.62     & 4.41$\pm$0.02 & 3.87$\pm$0.04  &  7.45$\pm$0.07 &  6.23$\pm$0.06 &  2.82$\pm$0.06  &  1.95$\pm$0.06  & 0.23$\pm$0.06  & 0.04   \\
G10.47    & 66.66$\pm$0.02 & 6.46$\pm$0.07  &  6.77$\pm$0.1   &  5.8$\pm$0.1   &  2.94$\pm$0.1  &  2.4$\pm$0.1 &  0.24$\pm$0.1  & 0.1   \\
G10.30    & 13.06$\pm$0.05 & 3.87$\pm$0.03  &  2.18$\pm$0.05 &  1.78$\pm$0.05 &  0.81$\pm$0.05 & 0.38$\pm$0.05 &  0.08$\pm$0.05  &  0.04   \\
G10.62    & --2.82$\pm$0.03 & 6.13$\pm$0.06  & 9.69$\pm$0.09  &  8.41$\pm$0.08  &  4.54$\pm$0.08  &  3.28$\pm$0.08  &  0.38$\pm$0.07  &  0.06  \\
G19.61    & 41.74$\pm$0.09 & 6.32$\pm$0.06  &  3.42$\pm$0.09 &  2.93$\pm$0.09
&  1.48$\pm$0.08 &  1.10$\pm$0.07 &  0.1$\pm$0.07   & 0.05    \\
G29.96    & 97.70$\pm$0.01 & 2.75$\pm$0.03  &  1.57$\pm$0.05 &  1.40$\pm$0.04 &  0.74$\pm$0.03 & 0.59$\pm$0.04 &  0.05$\pm$0.04 & 0.05   \\
G31.41    & 98.25$\pm$0.05 & 4.31$\pm$0.05 &  5.1$\pm$0.1 &  3.8$\pm$0.1  &     2.1$\pm$0.1  &  1.6$\pm$0.1  &  0.15$\pm$0.1 &  0.1  \\
G34.26    & 57.59$\pm$0.04 & 5.38$\pm$0.06  & 7.05$\pm$0.09 & 6.64$\pm$0.08 & 
 3.44$\pm$0.08 & 2.18$\pm$0.08 & 0.10$\pm$0.08 & 0.09   \\
G45.47    & 62.37$\pm$0.09 & 3.87$\pm$0.05  &  3.0$\pm$0.1  &  2.0$\pm$0.1  &    0.9$\pm$0.1  & 0.75$\pm$0.1  &  $\leq0.6$   &  0.1    \\
W51D      & 61.06$\pm$0.04 & 4.30$\pm$0.06  &  6.7$\pm$0.1  &  5.7$\pm$0.1  &  3.25$\pm$0.1 &  2.8$\pm$0.1  &  0.3$\pm$0.1  &  0.1   \\
IRAS\,20126& --3.70$\pm$0.07 & 2.18$\pm$0.02  & 1.84$\pm$0.04 & 1.52$\pm$0.04 & 0.65$\pm$0.05 & 0.39$\pm$0.04 &  $\leq0.2$  & 0.04   \\
\hline
\end{tabular}
\end{center}
\end{table*} 

\begin{table*}
\caption{Parameters of the \ace(6--5) lines observed with the FCRAO telescope.
G10.30 and W51D were not observed}
\begin{center}
\begin{tabular}{l|c c c c c c c }
\hline
Source  &  $v_{\rm LSR}$  &  $\Delta v_{1/2}$ & \multicolumn{4}{c}{$\int T_{\rm MB}{\rm d}v$ (K \kms)} &  $\sigma$ \\
\cline{4-7}
          &  (\kms)     &  (\kms)    & $K=0$ & 1 & 2 & 3 &  (K) \\
\hline
G5.89     & 8.79$\pm$0.05 & 3.46$\pm$0.03 & 6.7$\pm$0.2 & 5.6$\pm$0.2 &  3.0$\pm$0.2 & 2.0$\pm$0.2 & 0.2 \\
G9.62     & 4.2$\pm$0.1 & 4.2$\pm$0.1 &  2.9$\pm$0.1 & 2.6$\pm$0.1  & 1.2$\pm$0.1   &  0.6$\pm$0.1   & 0.1 \\
G10.47    & 66.9$\pm$0.2 & 5.4$\pm$0.2 & 2.2$\pm$0.2  & 1.9$\pm$0.1 & 0.9$\pm$0.1   &  0.7$\pm$0.1   & 0.09 \\
G10.62    & --2.9$\pm$0.1 & 5.3$\pm$0.1 &  4.3$\pm$0.2   & 3.2$\pm$0.2 & 1.5$\pm$0.2   & 1.1$\pm$0.2   & 0.1 \\
G19.61($^*$)  &  ---  &   ---   &  $\leq 3.0$  &  &  &  &  0.15 \\
G29.96    & 97.8$\pm$0.2 & 3.6$\pm$0.2 &  1.05$\pm$0.09 & 0.83$\pm$0.08 &  0.29$\pm$0.08 & 0.43$\pm$0.08 & 0.1 \\
G31.41    & 97.66$\pm$0.09 & 4.2$\pm$0.1 &  2.2$\pm$0.1 & 1.9$\pm$0.1 &  1.0$\pm$0.1 & 0.8$\pm$0.1 & 0.09 \\
G34.26    & 57.52$\pm$0.04 & 4.51$\pm$0.07 & 3.79$\pm$0.07 & 3.5$\pm$0.1   & 
 1.5$\pm$0.1  & 1.0$\pm$0.1   & 0.1 \\
G45.47    & 62.3$\pm$0.2 & 4.7$\pm$0.3 & 1.39$\pm$0.09 & 1.12$\pm$0.09 & 0.41$\pm$0.08 & $\leq 0.3$     & 0.07 \\
IRAS\,20126& --3.68$\pm$0.06 & 1.7$\pm$0.07 & 0.58$\pm$0.03 &  0.54$\pm$0.03 & 
 0.22$\pm$0.02 &  $\leq 0.2$  & 0.05 \\
\hline
\end{tabular}
\end{center}
 ($^*$) undetected source: the upper limits on $\int T_{\rm MB}$ have been
  obtained from $\sigma$ assuming a line FWHM equal to that measured with
  the IRAM 30-m telescope.
\end{table*}

\begin{table*} 
\begin{center}
\caption{Parameters of the \ace(8--7) lines observed with the IRAM 30-m telescope}
\begin{tabular}{l|c c c c c c c c}
\hline
Source  & $v_{\rm LSR}$ &  $\Delta v_{1/2}$ & \multicolumn{5}{c}{$\int T_{\rm MB}{\rm d}v$ (K \kms)} & $\sigma$ \\
\cline{4-8}
          & (\kms)  &  (\kms)     & $K=0$ & 1 & 2 & 3 & 4 & (K)  \\
\hline
G5.89     & 8.52$\pm$0.08 & 3.50$\pm$0.05  &  26.6$\pm$0.2 &  25.3$\pm$0.2 & 
  14.8$\pm$0.2  &  12.2$\pm$0.2  &  2.4$\pm$0.2  &  0.2  \\
G9.62     & 4.67$\pm$0.02 & 3.67$\pm$0.06  &  10.08$\pm$0.09 &  8.34$\pm$0.09 &  4.20$\pm$0.09  &  2.99$\pm$0.08  & 0.80$\pm$0.08  & 0.07  \\
G10.47    & 66.12$\pm$0.03 & 7.5$\pm$0.1  &  18.8$\pm$0.2   &  11.7$\pm$0.2   &  6.5$\pm$0.2   &  6.2$\pm$0.2   &  3.2$\pm$0.2  & 0.2  \\
G10.30    & 13.03$\pm$0.05 & 3.85$\pm$0.03  &  3.45$\pm$0.07 &  2.80$\pm$0.07 & 1.45$\pm$0.07 &  0.90$\pm$0.07 &  0.1$\pm$0.06   &  0.08  \\
G10.62    & --2.92$\pm$0.03 & 5.88$\pm$0.08  &  14.2$\pm$0.1  &  12.5$\pm$0.1  &   7.7$\pm$0.1  &  6.4$\pm$0.1  &  1.4$\pm$0.1  &  0.1  \\
G19.61    & 41.19$\pm$0.08 & 6.62$\pm$0.09  &  6.2$\pm$0.1 &  4.7$\pm$0.1 &  
  2.8$\pm$0.1 &  2.5$\pm$0.1 &  1.0$\pm$0.1    &  0.1   \\
G29.96    & 97.69$\pm$0.01 & 3.2$\pm$0.1  &  3.45$\pm$0.07 &  2.82$\pm$0.07 &    1.52$\pm$0.07 &  1.28$\pm$0.06 &  0.18$\pm$0.06 & 0.08   \\
G31.41    & 97.74$\pm$0.03 & 5.60$\pm$0.08 &  10.50$\pm$0.2 &  6.0$\pm$0.1  & 
  3.8$\pm$0.1  &  3.9$\pm$0.1  &  1.1$\pm$0.1 &  0.2   \\
G34.26    & 57.67$\pm$0.04 & 6.07$\pm$0.09  &  16.0$\pm$0.1 &  11.1$\pm$0.1 &    6.1$\pm$0.1 &  5.3$\pm$0.1 &  1.1$\pm$0.1 &  0.2  \\
G45.47    & 62.49$\pm$0.09 & 3.5$\pm$0.1  &  4.0$\pm$0.2  & 3.5$\pm$0.2  &   
  1.8$\pm$0.2  &  0.9$\pm$0.2 &  $\leq 0.7$      &   0.2   \\
W51D      & 60.84$\pm$0.04 & 5.22$\pm$0.1  &  15.0$\pm$0.2  &  10.7$\pm$0.2  &  7.3$\pm$0.2 & 6.3$\pm$0.2 &  1.7$\pm$0.2  &  0.2  \\
IRAS\,20126& --3.77$\pm$0.07 & 2.37$\pm$0.06  &  3.025$\pm$0.07 &  2.60$\pm$0.06 & 1.10$\pm$0.06 &  0.60$\pm$0.06 &  0.06$\pm$0.05 & 0.07  \\
\hline
\end{tabular}
\end{center}
\end{table*} 

\begin{table*} 
\begin{center}
\caption{Parameters of the \ace(13--12) lines observed with IRAM 30-m telescope}
\begin{tabular}{l|c c c c c c c c}
\hline
Source  & $v_{\rm LSR}$ &  $\Delta v_{1/2}$ & \multicolumn{5}{c}{$\int T_{\rm MB}{\rm d}v$ (K \kms)} & $\sigma$ \\
\cline{4-8}
          & (\kms)  &  (\kms)  & $K=0$ & 1 & 2 & 3 & 4 & (K)    \\
\hline
G5.89     & 8.56$\pm$0.06 & 3.92$\pm$0.06  &  29.0$\pm$0.3 &  28.6$\pm$0.3 &     18.6$\pm$0.3  &  18.0$\pm$0.3  &  4.0$\pm$0.3  &  0.4  \\
G9.62     & 4.78$\pm$0.05 & 4.16$\pm$0.06  &  6.7$\pm$0.2 &  6.0$\pm$0.2 &  
  4.2$\pm$0.2  &  3.6$\pm$0.2  & 0.5$\pm$0.2  &  0.2  \\
G10.47    & 66.55$\pm$0.1 & 8.0$\pm$0.1  &  15.0$\pm$0.4   &  16.0$\pm$0.2   &   9.6$\pm$0.2  &  9.5$\pm$0.1   &  3.1$\pm$0.1  &  0.4  \\
G10.30    & 12.79$\pm$0.08 & 3.47$\pm$0.09  &  2.6$\pm$0.1 &  2.4$\pm$0.1 &      1.4$\pm$0.1 &  1.0$\pm$0.1 &  0.2$\pm$0.1   &   0.2  \\
G10.62    & --2.74$\pm$0.05 & 6.1$\pm$0.1  &  14.6$\pm$0.3  &  14.5$\pm$0.3  &    9.6$\pm$0.3  &  9.6$\pm$0.2  &  2.2$\pm$0.2  & 0.3  \\
G19.61    & 40.8$\pm$0.1 & 9.1$\pm$0.2  &  5.7$\pm$0.3 &   10.0$\pm$0.3 &
  6.5$\pm$0.2 &  5.2$\pm$0.2 &  0.4$\pm$0.2   &  0.2  \\
G29.96    & 98.01$\pm$0.05 & 3.53$\pm$0.07  &  3.1$\pm$0.1 &  2.7$\pm$0.1 &   
  1.7$\pm$0.1 &  1.7$\pm$0.1 &  0.2$\pm$0.1 &  0.2  \\
G31.41    & 97.96$\pm$0.1 & 5.7$\pm$0.1 &   10.62$\pm$0.1 &  9.7$\pm$0.1  &      9.5$\pm$0.1  &  7.3$\pm$0.1  &  2.1$\pm$0.1 &  0.3  \\
G34.26    & 58.28$\pm$0.05 & 6.79$\pm$0.07  &  12.1$\pm$0.3 &  10.88$\pm$0.3 &   7.9$\pm$0.2 &  6.9$\pm$0.2 &  1.8$\pm$0.2 &  0.2  \\
G45.47    & 62.9$\pm$0.2 & 3.6$\pm$0.2  &   2.5$\pm$0.3  &  3.2$\pm$0.3  &    
  2.2$\pm$0.3  &  1.7$\pm$0.3 &  $\leq 1.1$    &  0.4   \\
W51D      & 60.59$\pm$0.06 & 4.99$\pm$0.07  &  16.0$\pm$0.2  &  15.9$\pm$0.2  &  10.0$\pm$0.1 &  9.8$\pm$0.3 &  1.6$\pm$0.3  &  0.4  \\
IRAS\,20126& --3.44$\pm$0.1 & 3.26$\pm$0.1  &  1.9$\pm$0.1 &  1.8$\pm$0.1 & 
  0.8$\pm$0.1 &  0.6$\pm$0.1 &   0.1$\pm$0.1   &  0.2  \\
\hline
\end{tabular}
\end{center}
\end{table*}

\subsection{Size}
\label{sthtr}

In order to assess that the \ace\ lines originate from the 1~pc clumps
known to exist around the UC \HII\ regions, we must estimate
the size of the \ace\ emitting region. The maps obtained with the QUARRY
multi-beam receiver on the FCRAO telescope are unsuitable to this
purpose, because emission was detected only towards the central position.
Consequently, only an upper limit to the source size could be obtained.
However, another estimate of the size can be derived from the comparison
between the \ace(6--5) line intensities measured with the IRAM 30-m and
FCRAO telescopes.
Assuming Gaussian intensity distributions for the \ace\ emitting clumps,
one finds that the main beam brightness temperatures, $T_{\rm MB}$,
measured with the two telescopes are related by the expression
\begin{equation}
T_{\rm MB}^{{\rm fcrao}}(\Theta_{\rm fcrao}^{2}+\Theta_{\rm s}^{2})=T_{\rm MB}^{\rm 30m}
(\Theta_{{\rm 30m}}^{2}+\Theta_{\rm s}^{2})
\end{equation}
where $\Theta_{\rm s}$ is the
FWHM of the source, and $\Theta_{\rm 30m}$ and 
$\Theta_{\rm fcrao}$ are the instrumental HPBWs. 

From this equation one obtains the expression for $\Theta_{\rm s}$:
\begin{equation}
\Theta_{\rm s}^2=\frac{\Theta_{\rm fcrao}^2-\Theta_{\rm 30m}^2(T_{\rm MB}^{{\rm 30m}}/T_{\rm MB}^{{\rm fcrao}})}{(T_{\rm MB}^{{\rm 30m}}/T_{\rm MB}^{{\rm fcrao}})-1}    \label{eth}
\end{equation}

\begin{figure}
\centerline{\psfig{figure=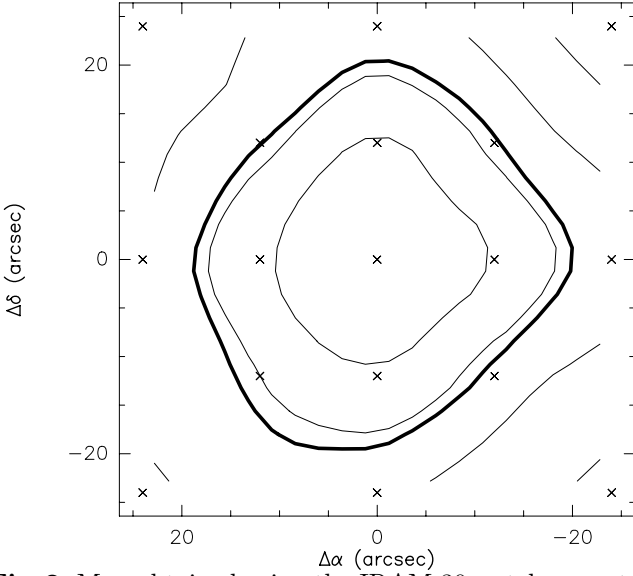,angle=-90,width=8.4cm}}
\caption{Map obtained
using the IRAM 30-m telescope
towards G31.41 by integrating the \ace(6--5)
 emission under the $K$=0 and $K$=1 lines. Contour levels range from 1~K~\kms\
 (corresponding to $3\sigma$) to 7~K~\kms\ in steps of 2~K~\kms. The
 thick line corresponds to the half maximum power contour}
\label{fmap65}
\end{figure}

\begin{figure}
\centerline{\psfig{figure=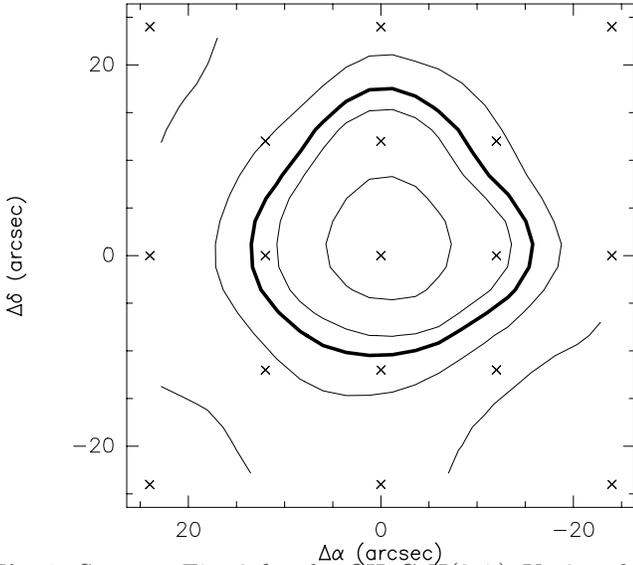,angle=-90,width=8.4cm}}
\caption{Same as Fig.~\ref{fmap65} for the \ace(8--7) $K$=0 and 1 lines.
 Contour levels range from 2 K \kms\ (corresponding to $3\sigma$) to 18
 K \kms\ in steps of 4 K \kms}
\label{fmap87}
\end{figure}

The values of $\Theta_{\rm s}$ are listed in Table~7 and the corresponding
linear diameters, $D$, are given in Table~\ref{mvir-mcd}. The errors have been
computed from Eq.~(\ref{eth}) with the usual propagation of the statistical
errors
and hence depend upon the errors on $T_{\rm MB}$, which will be
discussed in  Sect.~\ref{stempden}.
The mean values are $\Theta_{\rm s}\simeq30\arcsec$ and $D\simeq0.9$~pc.
We stress that in all cases
$\Theta_{\rm s}$ is consistent with the upper limits obtained from the
FCRAO maps. Also,
for G31.41 the value derived in this way  
can be compared with the direct estimate obtained from the  
emission map made with the
IRAM 30-m telescope. Figures~\ref{fmap65} and~\ref{fmap87}
show the maps of G31.41 obtained by
integrating respectively the (6--5) and (8--7) emission under the $K$=0
and 1 lines.
For a Gaussian source, one can compute the deconvolved
clump diameter from the observed FWHM.
We find $\Theta_{\rm s}=29\arcsec$ from the
\ace(6--5) map and $\Theta_{\rm s}=26\arcsec$ from \ace(8--7). These estimates
are consistent within the uncertainties with that obtained from Eq.~(\ref{eth}),
$\Theta_{\rm s}=32\arcsec\pm 8\arcsec$, thus confirming the
reliability of the method.
No size estimate can be derived from the
\ace(13--12) map because emission was detected only towards 
the central position: however, the upper limits obtained at offsets
of $\pm$12\arcsec\ from the centre
are consistent with the (13--12) line intensity measured towards the (0,0)
position and
the source diameter estimated above.

We can also compare our results with those derived from C$^{17}$O
(Hofner et al. 2000) and C$^{34}$S (Cesaroni et al. 1991):
in the former, the angular diameters are slightly greater ($\sim$20\%)
than those estimated by us, with the only exception of G10.47, which
is two times larger in C$^{17}$O; in the latter tracer, instead, the angular
sizes are $\sim$50\%
smaller with respect to \ace. We shall come back to this  
point in Sect.~\ref{sint}.

\subsection{Temperature and column density}
\label{stempden}

From the angular diameters obtained from Eq.~(\ref{eth}), we
derived the rotation temperature and total column density of the \ace\
molecules by means of the population diagram method
(see e.g. Hollis 1982, Olmi et al. 1993), which assumes
the gas to be in local thermodynamical equilibrium (LTE). Such an assumption
is believed to work very well for \ace\ (see Bergin et al. 1994)
due to its low dipole moment.
Under the additional assumption of optically thin emission, one can compute
the beam averaged column density $N_i$ of the upper level $i$ from the line
intensity:
\begin{equation}
\frac{N_i}{g_i}=\frac{3k\int T_{\rm MB}{\rm d}v}{8\pi^3\eta_\nu\mu^2\nu S}
\end{equation}
where $k$ is the Boltzmann constant, $\eta_{\nu}$ the beam filling factor, 
$\nu$ the line rest frequency, $S$ the line strength, $g_i$ the statistical 
weight, and $\mu=0.75$~Debye (Dubrulle et al. 1978) the dipole
moment of the \ace\ molecule.

\begin{figure}
\centerline{\psfig{figure=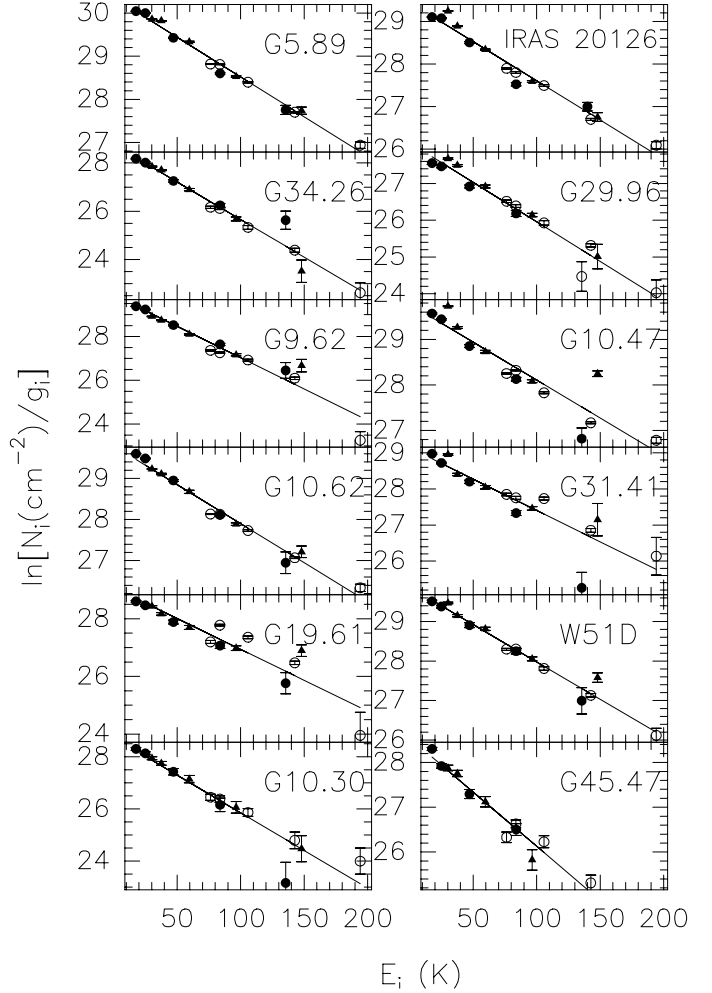,angle=-90,width=9.0cm}}
\caption{Boltzmann plots for all of the observed sources.
 Filled circles represent the \ace\ (6--5) data, triangles the
 (8--7), and empty circles the (13--12). The column densities have been obtained
 from the IRAM 30-m spectra. The straight lines represent least square fits
 to the data.
 The error bars correspond to the formal errors of the fitting procedure and do
 not consider the calibration error on $T_{\rm MB}$}
\label{grafico1}
\end{figure}

\begin{table}
\begin{center}
\caption{Angular diameters, rotation temperatures, and source averaged
 column densities of the \ace\ clumps. The errors are calculated 
 assuming a calibration error of $\sim10$$\%$ at 3~mm and $\sim20$$\%$ at 2~mm
 and 1.3~mm (see text)}
\label{tris1}
\begin{tabular}{lccc}
\hline 
Source  &  $\Theta_{\rm s}$  &  $T_{\rm rot}$  &  $N_{\rm tot}$   \\ 
          &  (arcsec)    &   (K)       &   (10$^{14}$cm$^{-2}$) \\
\hline
G5.89     &   31$\pm$8       &   48$\pm$10  &   20$\pm$1 \\
G9.62     &   28$\pm$7       &  34$\pm$6  &  13$\pm$1  \\
G10.47   &   22$\pm$5   &  46$\pm$8   &  27$\pm$1 \\
G10.30($^*$)&   25          &  34$\pm$7  &   5$\pm$1  \\
G10.62   &   27$\pm$8   &  46$\pm$10   &  21$\pm$1 \\
G19.61($^*$) &   30          &  56$\pm$7      &  11$\pm$3  \\
G29.96    &   45$\pm$9        &  48$\pm$8 &   4$\pm$1 \\
G31.41 &   32$\pm$8    &  53$\pm$8   &  15$\pm$2 \\
         &   29$\pm$5($^{**}$)    &  51$\pm$7    &  15$\pm$3 \\
         &   26$\pm$5($^{***}$)    &  48$\pm$7    &  16$\pm$3 \\
G34.26    &   40$\pm$9        &   46$\pm$9 &   16$\pm$2 \\
G45.47   &   35$\pm$8    &  39$\pm$7   &  5.0$\pm$0.3  \\
W51D($^*$)  &   23         &  47$\pm$10     &  15$\pm$3 \\
IRAS\,20126&   22$\pm$7         &   30$\pm$5  &   3.9$\pm$0.2 \\        
\hline 
\end{tabular}
\end{center}
($^*$) Sources for which the angular diameter could not be computed. For
G10.30 and W51D $T_{\rm rot}$ and $N_{\rm tot}$ have been computed assuming a
{\it linear} diameter equal to the mean value (0.9~pc) obtained from the
other sources. For G19.61, instead, we have assumed the diameter of 30\arcsec\
derived by Hofner et al. (2000) from the C$^{17}$O emission \\
($^{**}$) Angular diameter measured from the \ace(6--5) map of Fig.~\ref{fmap65} \\
($^{***}$) Angular diameter measured from the \ace(8--7) map of Fig.~\ref{fmap87}
\end{table}

In Table~\ref{tris1}
we give the rotational temperatures ($T_{\rm rot}$) and total 
\ace\ source column densities ($N_{\rm tot}$): these have been obtained from a
least square fit to the data, as shown in the ``Boltzmann plots''
in Fig.~\ref{grafico1}.
We note that each column density $N_i$ in Fig.~\ref{grafico1} includes the
correction for angular resolution effects, assuming gaussian profiles for
both the instrumental beam and the source. Therefore, $N_{\rm tot}$
corresponds to a mean value over the clumps.
The temperatures are distributed between 30 and 50~K and
the column densities are of order $10^{14}$--$10^{15}~{\rm cm}^{-2}$.
The goodness of the fits in Fig.~\ref{grafico1} confirms that the gas is very
close to LTE, which in turn implies that the rotational temperature derived in
this way is an excellent estimate of the kinetic temperature, $T_{\rm k}$,
of the molecular gas.

The errors quoted in Table~\ref{tris1} take into account calibration
uncertainties on $T_{\rm MB}$ of 10\% for the 3~mm lines and 20\% for the
others.
For $T_{\rm rot}$ and $N_{\rm tot}$, the uncertainties
have been estimated allowing for a variation of the values of $N_i$
by 10\% simultaneously for all of the \ace(6--5) $K$ lines and by 20\%
for the (8--7) and (13--12) lines: this takes into account the fact
that each sample of $J+1\rightarrow J$ transitions is affected by the same
calibration uncertainty. Then the errors have been taken equal to the
largest difference between the values of $T_{\rm rot}$ and $N_{\rm tot}$ thus
obtained and the nominal ones.

As previously explained, the population diagram method uses the
assumptions of
LTE and optically thin emission: let us now discuss these in some better
detail.

The LTE hypothesis is acceptable when the H$_2$ density in the clump is 
greater than the critical density (see e.g. Spitzer 1978).
The {\it maximum} critical density for the transitions observed by us
is $\sim10^5~{\rm cm}^{-3}$: we shall see
in the next section that the molecular hydrogen densities turn out to be
greater than this value, which supports our assumption of LTE
(and hence $T_{\rm rot}\simeq T_{\rm k}$) for the \ace\ molecules.

As for the optically thin assumption,
it is worth noting that the observed bandwidth covers most of the $K$
components of the CH$_3$$^{13}$CCH (6--5), (8--7), and (13--12) transitions.
One can thus estimate the optical depth from the intensity ratio between the
lines of the main species and those of the isotopomer. However, none of
the CH$_3$$^{13}$CCH lines is detected by us, so that only an upper limit
on the optical depth can be obtained: assuming an isotopic ratio
$^{12}$C/$^{13}$C=50 (see Wilson \& Rood 1994),
the minimum value turns out to be $\tau$=8, which
is insufficient to demonstrate that the CH$_3$C$_2$H lines are optically thin.
Nevertheless, the excellent agreement between the linear fit and the data
in the Boltzmann plots reinforces the assumption of optically thin line
emission: in fact, large optical depths are expected to affect mostly the low
excitation transitions and hence cause a flattening of the plots at low
energies, which is not seen in Fig.~\ref{grafico1}.

We will examine the
optical depth issue in some better detail in Sect.~\ref{stau}.

Another topic which may be discussed in relation to the fits in
Fig.~\ref{grafico1}, is the effect of temperature and density gradients
on the rotation diagrams: we delay this to Sects.~\ref{stgr} and~\ref{sdgr}.


\subsection{Mass and density}
\label{smass}

By means of the parameters derived above, we can estimate the mass of the clumps
and the corresponding virial mass needed for equilibrium. For both estimates,
spherical symmetry is assumed.

The mass of the clump, $M_{\rm CD}$, can be obtained from the column
density in Table~\ref{tris1}: for a spherical, isothermal clump one has
\begin{equation}
M_{\rm CD}=\frac{\pi}{4} \frac{\Theta_{\rm s}^{2}d^{2} N_{\rm tot}m_{\rm H_{2}}}{X_{\rm CH_{3}C_{2}H}}     \label{emcd}
\end{equation}
where $d$ is the distance to the source, $m_{\rm H_2}$ the mass of the H$_2$
molecule, $X_{\rm CH_3C_2H}$ the abundance of \ace\ relative to H$_2$,
and $N_{\rm tot}$ is the mean column density over the clump, as previously
explained.

The most uncertain value in Eq.~(\ref{emcd}) is $X_{\rm CH_3C_2H}$. Rather than
assuming a priori an arbitrary value for a parameter which is known to vary
by a large amount in the interstellar medium,
we have decided to obtain an estimate of it by equating Eq.~(\ref{emcd}) to
the mass, $M_{\rm cont}$, derived from continuum measurements of our sources
(the latter is given in Table~\ref{tcont}).
To this purpose, we used the (sub)millimeter continuum emission
observations made by various authors (Chini et al. 1986; Cesaroni et al. 1999;
Hatchell et al. 2000; Hunter et al. 2000) towards most of our sources.
Following Mezger et al. (1990), the mass was derived using these
continuum measurements and the dust temperatures quoted by the corresponding
authors; when the latter were not available, we used $T_{\rm rot}$ from
Table~\ref{tris1}.
The values of $X_{\rm CH_3C_2H}$ obtained by us range from
0.9~10$^{-9}$ to 5.7~10$^{-9}$ (see Table~\ref{tcont}) with a mean value
of $\sim2.2~10^{-9}$.
We have hence assumed $X_{\rm CH_3C_2H}=2~10^{-9}$, with an
uncertainty of a factor $\sim$2.5 and computed the $M_{\rm CD}$ from
Eq.~(\ref{emcd}) using this value {\it for all sources}. The results are
given in Table~\ref{mvir-mcd}.

It is worth pointing out that
the continuum measurements do not necessarily refer to the same region seen
in the \ace\ lines: however, in a few cases (G5.89, G9.62, G10.47, G31.41) maps
in the (sub)millimeter continuum are available, so that we could verify the
coincidence between the \ace\ and dust emitting regions ($\sim$30\arcsec).
For the remaining sources, the continuum observations were performed with an
instrumental HPBW greater than ours, so that the mass estimates in
Table~\ref{tcont} are to be taken as upper limits to the mass of the clumps
traced by \ace, which in turn imply lower limits on the \ace\ abundances;
however, the relatively small range of variation of $X_{\rm CH_3C_2H}$
in Table~\ref{tcont} suggests that our assumption of
$X_{\rm CH_3C_2H}=2~10^{-9}$ is likely to be a reliable approximation.

The virial mass for a homogeneous spherical clump,
neglecting contributions from magnetic field and surface pressure, can
be computed from the expression (see e.g. MacLaren et al. 1988)
\begin{equation}
M_{\rm vir}(M_\odot)=0.509\,d({\rm kpc})\,\Theta_{\rm s}({\rm arcsec})\,\Delta v_{1/2}^{2}({\rm km/s})   \label{emvir}
\end{equation}
where $\Delta v_{1/2}$ is the observed line FWHM.
In our case we took
$\Delta v_{1/2}$ from the \ace(6--5) lines, for the sake of consistency with
the derivation of $\Theta_{\rm s}$ which has also been obtained from the 
\ace(6--5) emission. The values of $M_{\rm vir}$ are given in
Table~\ref{mvir-mcd}.

In the same table
we also give the
H$_2$ volume densities estimated from $M_{\rm CD}$ and the clump diameters, $D$.
The uncertainties on these values have been calculated
by means of the propagation of statistical errors, but they do not take into
account distance uncertainties.
All masses are equal to a few $\sim 10^{3} M_{\odot}$,
with the sole exception of IRAS\,20126, for
which the mass is $\la10^2~M_\odot$. The $\rm H_{2}$ volume 
densities range from $10^{5}$ to $5~10^{6}$ cm$^{-3}$, greater than the 
maximum critical density of the transitions observed (see the discussion
in Sect.~\ref{sthtr}).


\subsection{Effects of inhomogeneity and optical depth}

In view of the findings that will be illustrated in Sect.~\ref{sdisc},
before proceeding further it is important to discuss the relevance
of optical depth effects, temperature gradients, and density gradients on
the mass (and hence density) estimates obtained so far.

\subsubsection{Optical depth}
\label{stau}
As explained in Sect.~\ref{stempden}, we believe that the bulk of the observed
\ace\ emission is optically thin. However, one cannot exclude the existence
of an optically thick region close to the centre of the clump, where the
density is presumably higher than average. Indeed, dense molecular cores
with diameters of $\sim$0.1~pc
are known to exist in most of our clumps.
Hence the question is how much of the clump mass is contained in the optically
thick region. One can calculate the size of the region over which $\tau\ge1$
for the \ace\ lines. For $N_{\rm tot}=2~10^{15}$~cm$^{-2}$,
$T=40$~K, $\Delta v_{1/2}=5$~\kms, and a density profile $n\propto R^{-2}$
one finds that the optical depth of e.g. the \ace(6--5) $K$=0 transition
along the line of sight passing through the centre of the clump
is equal to $\tau\simeq 0.1 (R_0/R-1)$, where $R_0$ is the radius of the
clump and $R$ the distance from the centre at which the optical depth $\tau$
is achieved. By posing $\tau=1$ one finds the radius of the optically
thick core, $R/R_0\simeq10$. In conclusion, the \ace\ gas
becomes optically thick at
a distance from the centre equal to $\sim$10\% of the clump radius:
this contains $\sim$10\% of the total mass. We conclude that optical
depth effects are negligible for the estimate of $M_{\rm CD}$.

Incidentally, it is worth noting that in case of large optical depths the
values of $N_{\rm tot}$ and $M_{\rm CD}$ are to be taken as lower limits.  On
the contrary, $M_{\rm vir}$ becomes an upper limit, because for $\tau\gg1$
line broadening occurs, which increases the value of $\Delta v_{1/2}$ used in
Eq.~(\ref{emvir}). As a consequence, the ratio $M_{\rm CD}/M_{\rm vir}$ which
one can derive from Table~\ref{mvir-mcd} has to be regarded as a lower limit.
This fact reinforces the conclusion attained later in Sect.~\ref{sstab}.

\subsubsection{Temperature gradient}
\label{stgr}
The hypotheses under which the method of the population diagrams in
Sect.~\ref{stempden} is applied are that the gas is optically thin, in LTE,
and isothermal. The first two assumptions have already been discussed;
as for the latter, Mauersberger et al. (1988) have treated the case of a
spherically symmetric cloud with temperature and density having power-law
dependence on the distance from the centre. These authors demonstrate that in
this case a linear correlation should hold between $\ln N_i$ and $\ln E_i$,
unlike the isothermal case where the correlation is expected with $E_i$.
We stress that the different dependence on $E_i$ is caused {\it only} by
the temperature gradient: in fact, for an isothermal cloud the usual linear
relation between $\ln N_i$ and $E_i$ is recovered, no matter what the density
gradient is.

\begin{figure}
\centerline{\psfig{figure=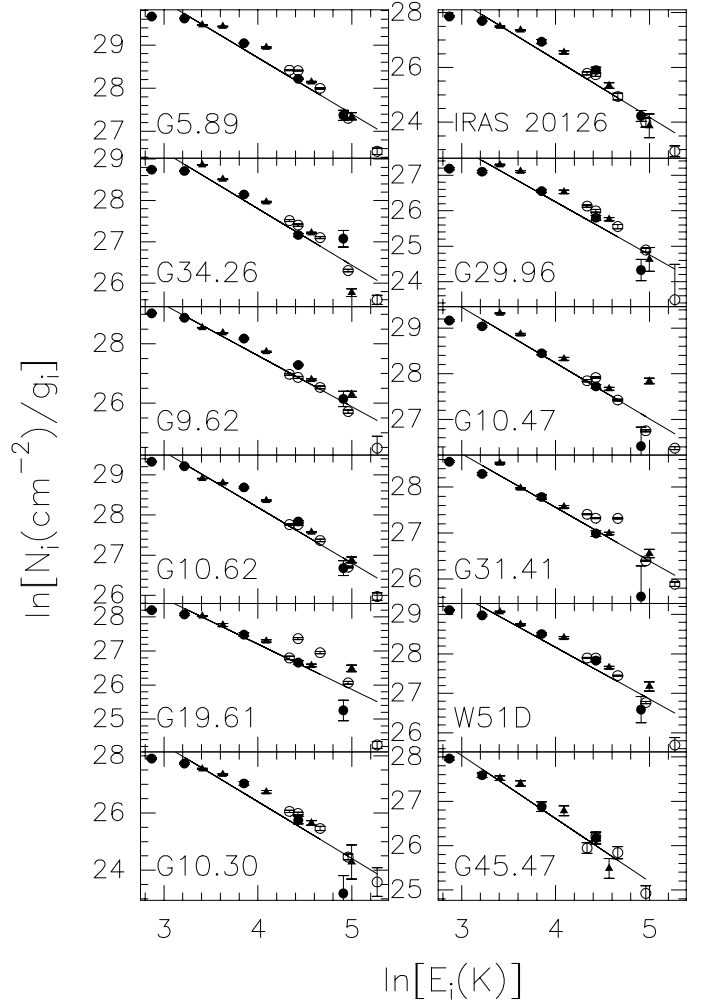,angle=-90,width=9.0cm}}
\caption{
Same as Fig.~\ref{grafico1} but with logarithmic scale on the
 abscissa
}
\label{fmodbol}
\end{figure}

Following the example of Mauersberger et al. (1988), in Fig.~\ref{fmodbol} we
present ``modified'' population diagrams for our sources, plotting $\ln N_i$
versus $\ln E_i$. Clearly, the linear fits
to the points are much less satisfactory than those in Fig.~\ref{grafico1}.
We conclude that the \ace\ emission arises from a region in which the
temperature profile must be very shallow, hence making $T$=const. a suitable
approximation.

\subsubsection{Density gradient}
\label{sdgr}
For an isothermal and optically thin cloud, the
source averaged column density cannot depend on the way the gas is distributed
inside the cloud: hence, density gradients leave the estimates of
$N_{\rm tot}$ and $M_{\rm CD}$ unaffected. On the other hand, the virial
mass depends on the density profile, as demonstrated e.g. by
MacLaren et al. (1988). For a power-law density distribution of the type
$n_{\rm H_2}\propto R^p$, the virial mass obtained from Eq.~(\ref{emvir})
must be multiplied by the factor $\frac{3}{5}\frac{5+2p}{3+p}$, which is
$\le$1 for $p\le0$. Therefore, the values of $M_{\rm vir}$ of
Table~\ref{mvir-mcd} derived for a homogeneous cloud are to be taken as upper
limits.

In conclusion, the ratio $M_{\rm CD}/M_{\rm vir}$ derived by us
represents an underestimate if the constant density assumption is released
in the computation of $M_{\rm vir}$.
This result is important in view of the discussion in Sect.~\ref{sstab}.


\begin{table}
\begin{center}
\caption{Mass and \ace\ abundance estimates from the continuum millimeter
 emission}
\label{tcont}
\begin{tabular}{lccccc}
\hline
Source  & $d^{(a)}$ & $M_{\rm cont}$ & $X_{\rm CH_3C_2H}$ & Ref.  \\
          & (kpc) & ($M_{\odot}$) & ($\times 10^{-9}$)  &   \\ \hline
G5.89     & 4.0 & 1500 &  5.5 & 1 \\
G9.62      & 5.7 & 1600 &  5.7 &  1 \\
G10.47     & 5.8 & 12700 &  0.9 & 2 \\
G10.30    & 6.0 & --- & ---  &  --- \\
G10.62     & 6.0 & 8700 &  1.7 & 3  \\
G19.61    & 3.5 & 2200 &  1.4 & 3 \\
G29.96    & 7.4 & 11200 &  1.0 & \\
G31.41    & 7.9 & 7300 &  3.5 & 2 \\
G34.26    & 4.0 & 9700 &  1.1 & 3 \\
G45.47     & 8.3 & 12900 &  0.9  & 3 \\
W51D      & 8.0 & --- & ---  & --- \\
IRAS\,20126 & 1.7 & 110 &  1.3 & 4 \\
\hline
\end{tabular}
\vskip 0.1cm
 Reference codes:
 1=Hunter et al. (2000); 2=Hatchell et al. (2000);
 3=Chini et al. (1987); 4=Cesaroni et al. (1999)\\
\end{center}
$^{(a)}$~distances are from Wood \& Churchwell (1989a) for G10.30 and from
 Kurtz et al. (2000) for the other sources \\
\end{table}

\section{Discussion}
\label{sdisc}

In the previous section, we have demonstrated that the values of the physical
parameters derived by us are robust. In particular, we have shown that:
(i)~optical depth effects can be neglected for the \ace\ lines;
(ii)~temperature gradients, if present, must be very ``flat'' in the \ace\
 emitting region; and
(iii)~density gradients have no effect on the estimate of $M_{\rm CD}$
 and the corresponding H$_2$ density.
In the following, we shall use these physical parameters to analyse the
stability and structure of the clumps.


\subsection{Stability of the clumps}
\label{sstab}

From Table~\ref{mvir-mcd} one can see that the mean ratio between $M_{\rm CD}$
and $M_{\rm vir}$ is $\sim$3.0$\pm$1.7, being $\sim$1 only in the cases of
G19.61 and IRAS\,20126. Although such a ratio is only marginally $>$1,
it is such for all of the sources but two of them: this result
seems too systematic
to be due only to random errors on the quantities of interest.
The question is whether the ratio $M_{\rm CD}/M_{\rm vir}$ can be reduced
to unity by any means. Let us examine a few possibilities.


\begin{figure}
\centerline{\psfig{figure=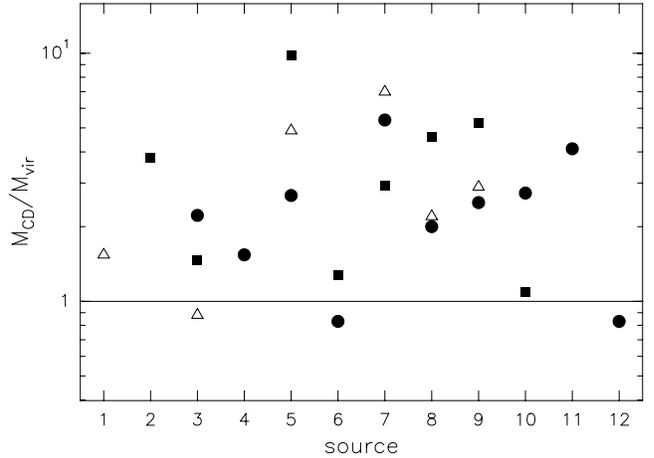,angle=-90,width=8.5cm}}
\caption{Plot of the ratio between the mass estimated from the gas column
 density and that derived assuming virial equilibrium for the clumps observed
 in this study. Full circles indicate values computed from \ace\ (this study),
 empty triangles those derived from C$^{34}$S (Cesaroni et al. 1991), and
 filled squares those obtained from C$^{17}$O (Hofner et al. 2000).
 The straight line corresponds to $M_{\rm CD}=M_{\rm vir}$}
\label{fmhof}
\end{figure}

As already discussed in Sects.~\ref{stau} and~\ref{sdgr}, optical depth
effects and density gradients can only increase the ratio
$M_{\rm CD}/M_{\rm vir}$, as they make $M_{\rm CD}$ bigger and
$M_{\rm vir}$ smaller.

Another possibility is that the \ace\ abundance is underestimated, which would
cause an overestimate of $M_{\rm CD}$. In principle this cannot be excluded,
given the large variations of molecular abundances in different objects (see
e.g. Irvine et al. 1987); however, our estimate of $X_{\rm CH_3C_2H}$ (see
Sect. \ref{smass}) is ``direct'', as it makes use of the continuum emission
from the regions of interest and it is hence less prone to errors which instead
affect values based on chemical models. Further evidence for
$M_{\rm CD}$ being greater than $M_{\rm vir}$ comes from the studies of
Cesaroni et al. (1991) and Hofner et al. (2000) who mapped some of our clumps
respectively in the C$^{34}$S and C$^{17}$O lines: as one can see in
Fig.~\ref{fmhof}, in all sources the ratio $M_{\rm CD}/M_{\rm vir}$ is
basically $>$1, independently of the tracer. It is
worth stressing that the values of $M_{\rm CD}$ obtained by Hofner et
al. (2000) from C$^{17}$O are less affected by uncertainties on the
abundance, which is much better established for C$^{17}$O than  
for \ace.

In conclusion, we favour the hypothesis that the
discrepancy between $M_{\rm CD}$ and $M_{\rm vir}$ is real.
In this case $M_{\rm CD}/M_{\rm vir}>1$ seems to indicate that
the clumps are unstable against gravitational collapse.

\begin{figure}
\centerline{\psfig{figure=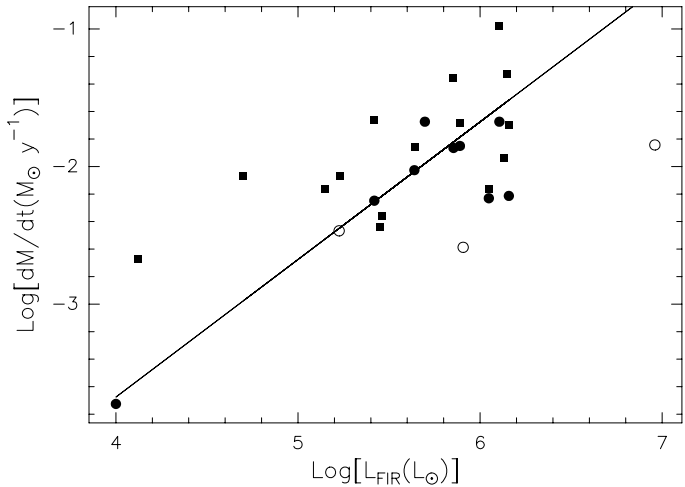,angle=-90,width=9cm}}
\caption{Plot of the mass accretion rate versus the source bolometric
 luminosity estimated from the IRAS fluxes. Circles and squares indicate
 estimates obtained respectively from \ace\ (this study) and
 C$^{17}$O (Hofner et al. 2000). Empty circles indicate the
 sources for which it was not possible to estimate the size of the \ace\
 clump (see Table~\ref{tris1}). The straight line is the best
 fit assuming proportionality between mass accretion rate and luminosity and
 corresponds to
 $\dot{M}_{\rm acc}(M_\odot{\rm yr^{-1}})=2.1~10^{-8}L_{\rm FIR}(L_\odot)$}
%
\label{fratlum}
\end{figure}

If thermal pressure and turbulence are the only means of support against
gravitational collapse, indeed $M_{\rm CD}>M_{\rm vir}$ implies collapse of
the clumps on a free-fall time: for densities of order $10^6$~cm$^{-3}$
the latter is equal to $\sim$$10^4$~yr, an order of magnitude less than
the estimated lifetime of the UC~\HII\ regions embedded in the clumps
(Wood \& Churchwell 1989b).
This suggests that magnetic fields might play a role in stabilising
the cloud against gravitational collapse.
From the virial theorem one can estimate the strength of the magnetic field,
$B$, required to this purpose: following
McKee et al. (1993) and neglecting the surface pressure term, the expression
for virial equilibrium can be written as
\begin{equation}
3 M \frac{\Delta v_{1/2}^2}{8\ln 2} + \frac{b}{3} B^2 R^3 =
 \frac{3a}{5} \frac{GM^2}{R}
\end{equation}
where $a$ and $b$ are two factors of order unity that take into account the
cloud geometry and the topology of the magnetic field. The equilibrium value
of $B^2$ is hence given by
\begin{equation}
B^2 = \frac{9}{5} \frac{a}{b} \frac{G M}{R^4}
      \left(M-\frac{5}{8\ln 2} \frac{R \Delta v_{1/2}^2}{a G}\right).
\end{equation}
Assuming $a$=1.17 and $b$=0.3 (McKee et al. 1993) and
posing $M=M_{\rm CD}$ we can compute the values of $B$ required to support
our clumps: these are a few mG. Such values
are comparable to those obtained by Lai et al. (2001)
who measured the magnetic field in molecular clouds with
densities similar to those of our clumps using interferometric polarisation
maps.

We thus propose a scenario in which the clumps are marginally stable due to
the contribution of the magnetic field. In this case the clump will
contract on the timescale $t_{\rm AD}$ determined by ambipolar diffusion.
This can be related to the free-fall time, $t_{\rm ff}$, by means of
Eq.~(51) of McKee et al. (1993): $t_{\rm AD}\simeq8.5\,t_{\rm ff}$, thus
obtaining $t_{\rm AD}({\rm yr})=2.9~10^8~[n_{\rm H_2}({\rm cm^{-3}})]^{-0.5}$.
The mass accretion rate onto the embedded stars can be hence computed
from the ratio $\dot{M}_{\rm acc}=M_{\rm CD}/t_{\rm AD}$ and plotted as
a function of the
source luminosity, as done in Fig.~\ref{fratlum}: in this figure we have also
reported the estimates obtained from the C$^{17}$O data of Hofner et al. (2000).
The accretion
rate increases with source luminosity: this is consistent with the
finding that the star formation
rate in a cloud is roughly proportional to the bolometric luminosity
according to the expression
$\dot{M}_\ast(M_\odot{\rm yr^{-1}})\simeq4~10^{-10}\,L(L_\odot)$
(Plume et al. 1997). Therefore,
it is reasonable to expect that also $\dot{M}_{\rm acc}$ satisfies
a similar relationship. We hence fitted the data in Fig.~\ref{fratlum}
assuming $\dot{M}_{\rm acc}\propto L_{\rm FIR}$, thus obtaining
$\dot{M}_{\rm acc}/L_{\rm FIR}=2.1~10^{-8}~M_\odot{\rm yr^{-1}}L_\odot^{-1}$.
The ratio $\dot{M}_\ast/\dot{M}_{\rm acc}\simeq1.9\%$ is the star formation
efficiency in the clumps. On the other hand, a lower limit to the same value
can be obtained from the ratio between the mass of the early-type star
ionising the embedded UC \HII\ region and the mass of the corresponding
clump: this is approximately $30~M_\odot/5000~M_\odot\simeq0.6\%$, consistent
with the previous estimate.
It is also worth noting that the mass accretion rates are of the order of
$\sim$10$^{-2}~M_\odot\,{\rm yr}^{-1}$: values such high could support
theories which predict that
high-mass stars can form through accretion (Behrend \& Maeder 2001).

In conclusion, although other explanations are possible, we cannot
exclude the possibility that the molecular clumps traced by the
\ace\ transitions are on the edge of gravitational collapse.
In support of this hypothesis we note that evidence for
infall in some of our clumps has been found by other authors
on the basis of high angular resolution observations in various
molecular tracers (Hofner et al. 1999; Maxia et al. 2001). In
Sect.~\ref{sdens} we will demonstrate that also the density structure of
the clumps favours the scenario presented above.

\begin{table*}
\begin{center}
\caption{Mass and H$_2$ density estimates}
\label{mvir-mcd}
\begin{tabular}{lcccc}
\hline
Source  &  $D$ & $M_{\rm vir}$ &  $M_{\rm CD}^{(a)}$ & $n_{\rm H_2}^{\rm CD}$ \\
        &  (pc) & ($M_{\odot}$) & ($M_{\odot}$) & (10$^{5}$cm$^{-3}$) \\
\hline
G5.89     &  0.6$\pm$0.2 & 1800$\pm$300  & 4500$\pm$1200 & 8$\pm$2 \\
G9.62     &  0.8$\pm$0.2 & 1000$\pm$400  & 5000$\pm$1000 & 5$\pm$1 \\
G10.47    &  0.6$\pm$0.1 & 2700$\pm$1000 & 6000$\pm$2000 & 10$\pm$3  \\
G10.30    &  0.8         & 1300$\pm$400  & 2000$\pm$1000 & 20$\pm$10\\
G10.62    &  0.8$\pm$0.2 & 3000$\pm$1000 & 8000$\pm$2000 & 2.5$\pm$0.8  \\
G19.61    &  0.5         & 1800$\pm$500  & 1500$\pm$500  & 7$\pm$2 \\
G29.96    &  1.6$\pm$0.3 & 1300$\pm$400  & 7000$\pm$2000 & 0.7$\pm$0.2\\
G31.41    &  1.2$\pm$0.3 & 2500$\pm$600  & 5000$\pm$2000 & 3$\pm$1 \\
G34.26    &  0.8$\pm$0.2 & 2400$\pm$700  & 6000$\pm$1500 & 6$\pm$2 \\
G45.47    &  1.4$\pm$0.3 & 2200$\pm$800  & 6000$\pm$2000 & 0.9$\pm$0.3 \\
W51D      &  0.9         & 1700$\pm$600  & 7000$\pm$3000 & 50$\pm$20 \\
IRAS\,20126 &  0.18$\pm$0.06 & 90$\pm$30 & 75$\pm$20 & 5$\pm$1 \\
\hline
\end{tabular}
\end{center}
$^{(a)}$~calculated assuming a \ace\ abundance of $2~10^{-9}$
\end{table*}

\subsection{Internal structure of the clumps}
\label{sint}

We have found that the clumps seen in the \ace\ lines have mean diameters of
0.9~pc, kinetic temperatures of $\sim$40~K and H$_2$ volume densities of
$10^5$--$10^6$~cm$^{-3}$. It is interesting to compare these values
with those derived from other tracers.

For the same quantities,
Hofner et al. (2000) find slightly different mean values from their
$\rm C^{17}O$ observations, namely diameters of $\sim$1.1~pc,
kinetic temperatures of $\sim$20~K and H$_2$ volume densities of
$\sim$10$^4$~cm$^{-3}$. This means that the $\rm C^{17}O$ emission arises
from larger, colder, and less dense regions than \ace.
This is not surprising as the $\rm C^{17}O$ is an order of magnitude more
abundant than \ace\ and hence likely to trace lower density gas.

On the other hand, the C$^{34}$S observations of Cesaroni et al. (1991) and
Olmi \& Cesaroni (1999) provide greater densities and smaller diameters than
our \ace\ measurements, namely $\sim$$10^6$~cm$^{-3}$ and $\sim$0.4~pc.

Finally, Kurtz et al. (2000) compiled a list of the properties of hot
molecular cores (see their Table~1): such cores are present in most
of our sources and represent the innermost, densest parts of the clumps seen
in \ace, with typical temperatures of $\sim$100~K and diameters $\la$0.1~pc.

In the following we make use of all these information to derive the
distribution of the physical parameters in the clumps.

\subsubsection{Temperature structure}
\label{stemp}

\begin{figure}
\centerline{\psfig{figure=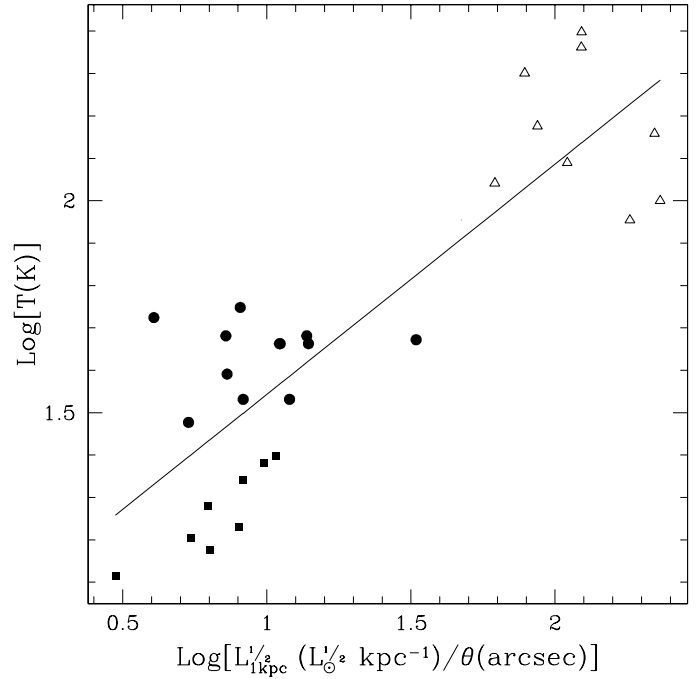,angle=0,width=9cm}}
\caption{Plot of the kinetic temperature versus the quantity
 $L_{\rm 1kpc}/\theta$, where $L_{\rm 1kpc}$ is the source luminosity
 scaled to the distance of 1~kpc and $\theta$ the angular radius
 at which the temperature has been measured. Filled circles and squares
 indicate respectively estimates based on \ace\ (this study) and
 C$^{17}$O (Hofner et al. 2000), while empty triangles correspond to
 the values of Table~1 of Kurtz et al. (2000). The straight line represents
 a least square fit to the data, with slope $0.54\pm0.06$ and intercepta
 $1\pm0.09$
 }
\label{ftlum}
\end{figure}

If we assume that the gas of the clump is heated by an embedded high-mass
star,
the gas temperature, $T$, at a distance $R=\theta\,d$ from the
star is given by the expression (see e.g. Plume et al. 1997 and references
therein)
\begin{equation}
T = C \left(\frac{L^\frac{1}{2}}{\theta\,d}\right)^{-q}   \label{etemp}
\end{equation}
with $d$ distance to the source, $L$ luminosity of the embedded star, and
$\theta$ angular distance from the star.
In order to verify whether such a relation
holds for our clumps and to compute the values of $C$ and $q$,
in Fig.~\ref{ftlum} we plot $\Log(T)$ versus
$\Log[L_{\rm FIR}^{1/2}/(\theta\,d)]$: note that both quantities are distance
independent. The distribution can be fitted with a straight line
corresponding to $q=-0.54\pm0.06$ and $C=10\pm2$, when $T$ is expressed in K,
$L_{\rm FIR}$ in $L_\odot$, $d$ in kpc, and $\theta$ in arcsec.

\begin{figure}
\centerline{\psfig{figure=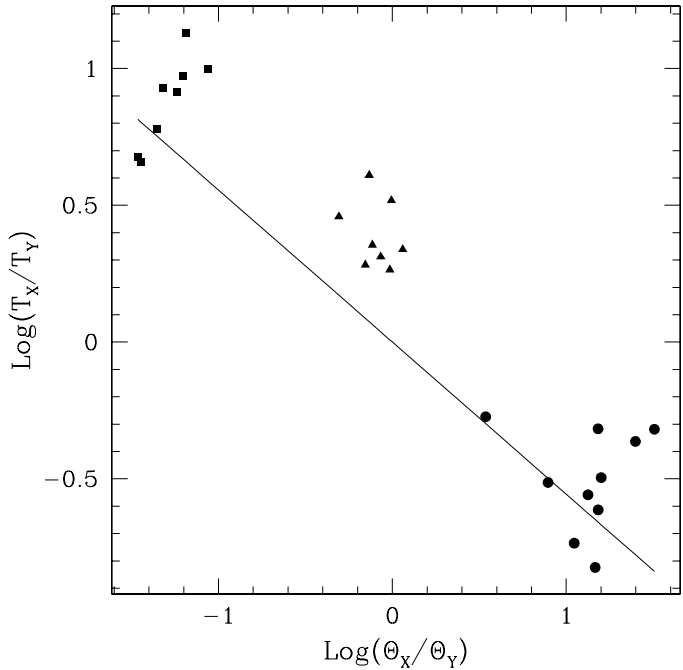,angle=0,width=9cm}}
\caption{Plot of the temperature ratio $T_{\rm X}/T_{\rm Y}$ versus the
 angular diameter ratio $\Theta_{\rm X}/\Theta_{\rm Y}$, where X and Y
 indicate two estimates among those derived from \ace, C$^{17}$O
 (Hofner et al. 2000), and those of Table~1 of Kurtz et al. (2000).
 Circles correspond to ratios between the temperatures quoted by Kurtz et al.
 and those obtained from \ace, squares to ratios between estimates from
 Kurtz et al. and from C$^{17}$O, and triangles to ratios between
 estimates from \ace\ and C$^{17}$O. The straight line represents a least
 square fit of the type $T\propto \theta^q$, with $q=-0.56\pm0.06$. The 
 correlation coefficient is --0.93
 }
\label{ftemp}
\end{figure}

Equation~(\ref{etemp}) implies that for a given source $T \propto \theta^q$.
Therefore, one can also
derive $q$ from the ratio between the
temperatures measured in two different tracers and the
ratio between the corresponding diameters, which satisfy the relation
\begin{equation}
\frac{T_{\rm X}}{T_{\rm Y}}=\left(\frac{R_{\rm X}}{R_{\rm Y}}\right)^q 
    \label{etq}
\end{equation}
with X,Y indicating that the value has been obtained from \ace, C$^{17}$O
(Hofner et al. 2000), or taken from Table~1 of Kurtz et al. (2000).
Under the assumption that $q$ is exactly the same for all sources,
Eq.~(\ref{etq}) is totally source independent
and one can plot all possible ratios
{\it for all sources} in the same figure.
The advantage of this approach with respect to that previously used is that
we do not need any luminosity estimate.
The result is shown in Fig.~\ref{ftemp}:
the slope in this case is $q=-0.56\pm0.06$, consistent with the previous
estimate.

We conclude that the temperature in the clumps is approximately described by
the law $T \propto R^{-0.55}$.
Such a law is steeper than expected on the basis of the discussion of
Sect.~\ref{stgr}, where we conclude that no temperature gradient is seen
in the gas traced by the \ace\ emission. However, in order to reveal this
gradient we had to compare the temperature measured on the
large scale (in the \ace\ and C$^{17}$O emission) with that on
a very small scale (in CH$_3$CN). This means that while $T \propto R^{-0.55}$
is a reasonable description of the ``global'' temperature variation
from $\sim$0.1~pc to $\sim$1~pc, it cannot be applied rigorously at
all radii in the clump. In particular, the temperature profile is likely
to be shallow in the outer region (for the reason discussed in Sect.~\ref{stgr})
and steeper close to the center.

Note also that $q$=--0.55 is significantly greater than
the slope expected for an optically thin clump heated by an embedded energy
source. In fact, in this case for a dust absorption coefficient
$\propto$$\nu^\beta$ one finds
$T \propto R^{-2/(4+\beta)}$: the steepest slope is obtained for
$\beta=1$ and is equal to --0.4.
A plausible explanation for $q$ being less than --0.4
is that part of the clump is optically thick. Indeed,
in some of our objects, on scales 10 times smaller than the clumps,
Cesaroni et al. (1998) found optically thick, hot cores
with temperature profiles
$T \propto R^{-0.75}$. These cores are just the regions
which the temperature estimates by Kurtz et al. (2000) used in our plots
refer to. We thus conclude that
one may read the slope derived by us as a ``compromise'' between a flatter
temperature profile in the optically thin clump --
traced by C$^{17}$O and \ace\ -- and the steeper profile ($T\propto R^{-0.75}$)
in the hot core -- measured e.g. in NH$_3$ or CH$_3$CN.

\subsubsection{Density structure}
\label{sdens}

For star forming clumps, theoretical models predict density profiles of the
type $n\propto R^p$. In our case, it is difficult to test this
prediction source by source, as only few measurements for each source at
different radii are available, namely those from \ace, C$^{34}$S, C$^{17}$O,
and the density estimates obtained from Table~1 of Kurtz et al. (2000).
However, the ratio between the H$_2$ densities measured with different tracers
is equal to
\begin{equation}
\frac{n_{\rm X}}{n_{\rm Y}}=\left(\frac{R_{\rm X}}{R_{\rm Y}}\right)^p
 \label{edens}
\end{equation}
with $n_{\rm X,Y}$ density measured with tracer X,Y=\ace, C$^{17}$O, C$^{34}$S,
and mm continuum (from Kurtz et al. 2000),
and $R_{\rm X,Y}$ radius of the spherical
region traced by X,Y. 
One can demonstrate that if Eq.~(\ref{edens}) holds for the volume density at
radius $R$, then it must hold also for the mean density inside $R$, {\it and
vice versa}. This result is important because it makes possible to use the
mean densities given in Table~\ref{mvir-mcd} to derive an estimate of $p$.

\begin{figure}
\centerline{\psfig{figure=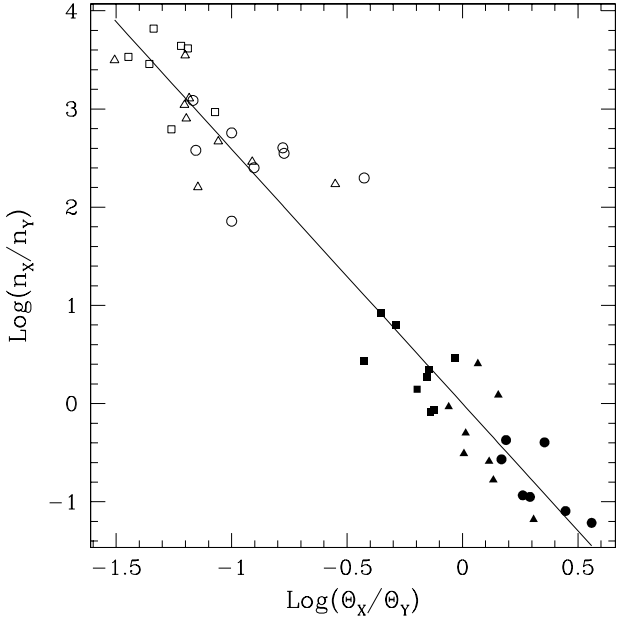,angle=0,width=9cm}}
\caption{Same as Fig.~\ref{ftemp} for the H$_2$ density. Filled symbols:
 circles correspond to ratios between densities from C$^{17}$O and C$^{34}$S,
 squares to ratios between densities from C$^{34}$S and \ace, and triangles
 to ratios between densities from C$^{17}$O and \ace. Empty symbols:
 circles, triangles, and squares correspond respectively to ratios between
 densities estimated from Table~1 of Kurtz et al. (2000) and those
 derived from C$^{17}$O, \ace, and C$^{34}$S. The straight line
 represents a least square fit of the type $n\propto \theta^p$, with
 $p=-2.59\pm 0.07$. The correlation coefficient is --0.97
 }
\label{fdens}
\end{figure}

As already done for the temperature profile,
we assume that $p$ is the same for all sources. Under this
hypothesis,
one can plot
$\Log(n_{\rm X}/n_{\rm Y})$ versus $\Log(\Theta_{\rm X}/\Theta_{\rm Y})$
for all possible ratios {\it and all sources}:
this is shown in Fig.~\ref{fdens}.
The correlation is excellent (the correlation coefficient is --0.97) and well
reproduced by a linear fit
with $p=-2.59\pm 0.07$.
 This result is consistent with the
density profiles predicted by star formation models
(Shu et al. 1987; Li 1999)
and supports the idea discussed in Sect.~\ref{sstab} that our clumps are
marginally stable.
Noticeably,
in W3(H$_2$O) -- an object of the same type as those observed by us -- an even
steeper density law has been derived by Mauersberger et al. (1988): they
find indirect evidence for $p$ ranging between --3 and --2.8. This is
in reasonable agreement with our finding.

It is worth noting that the previous density law applies to both the core
(empty points in Fig.~\ref{fdens}) and the surrounding clump (filled
points) independently. This result is not entirely consistent with
the findings of Hatchell et al. (2000), who fit their continuum maps
of similar clumps around UC~\HII\ regions with shallower density
profiles. However, their fits are model dependent, whereas our results have
been obtained under relatively simple assumptions, which in our opinion
make them more reliable.

On the other hand, we know that massive stars have already formed
inside the clumps, which means that gravitational collapse is going on
in the central regions. According to Shu et al. (1987) this occurs within a
radius
$R_2({\rm pc})=1.72~10^{-3}~M_\ast(M_\odot)\,[\Delta v_{1/2}({\rm km\,s^{-1}})]^{-2}$,
where $M_\ast$ is the mass of the star created by the collapse: inside
this radius the density should be $\propto$$R^{-3/2}$. In our case,
the maximum value of $R_2$ is obtained for $M_\ast\simeq60~M_\odot$ and
$\Delta v_{1/2}\simeq2$~\kms, which give $R_2\simeq 0.026$~pc, much less
than the minimum radius measured in any of the tracers used in Fig.~\ref{fdens}:
this confirms that the density profile measured with such tracers can be
$\propto$$R^{-2}$, close to our result.

It is worth noting that $n\propto R^{-2}$ holds for an {\it isothermal sphere},
which seems to contradict the result $T\propto R^{-0.55}$ found in
Sect.~\ref{stemp}. However, the density gradient is established on dynamical
timescales, much longer than the radiative timescale needed for the newly
born high-mass stars to heat up the clumps. As a result, the density profile
is a ``remnant'' of the primeval structure of the clump, whereas the
temperature profile reflects the present heating by the embedded stars.

\subsubsection{Velocity structure}

Given the low angular resolution of our observations,
the only possibility to investigate the velocity field in the clumps
is to study the variation of the line shape as a function of
the corresponding excitation energy: transitions from higher energy
levels are expected to arise from inner regions since the temperature
increases towards the center of the clumps, as previously discussed.
This makes possible to sample the clumps at different radii by means of the
different $K$ lines of \ace.

\begin{figure}
\centerline{\psfig{figure=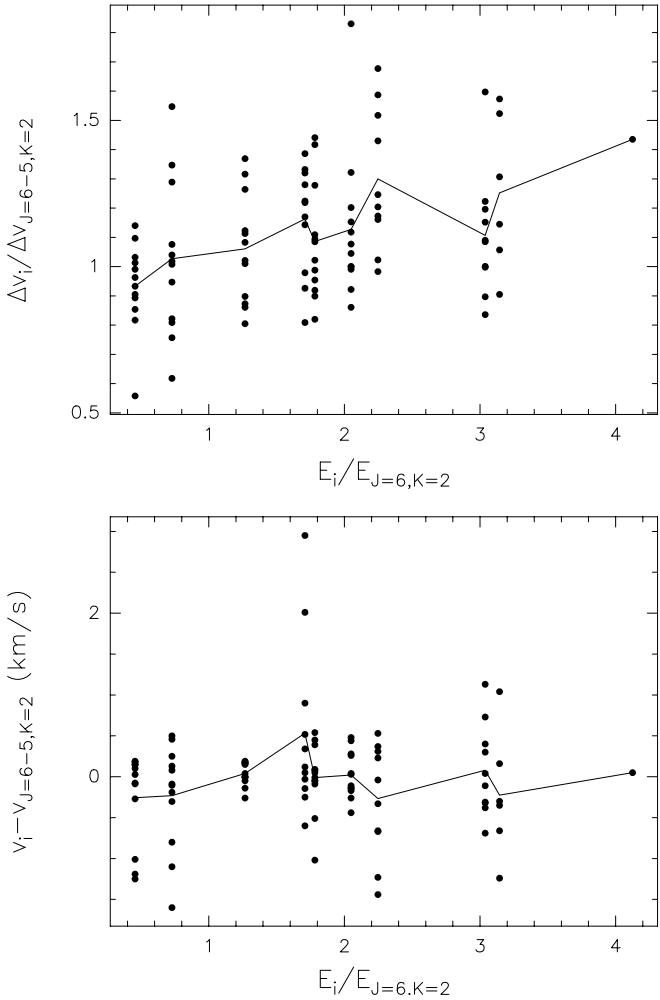,angle=0,width=8.8cm}}
\caption{Top panel: plot of the ratio between the FWHM of a \ace\ $K$ line 
and the FWHM
of the $(6-5)$ $K$=2 line, versus the corresponding excitation energy ratio.
The line connects the mean values of the ratios obtained for each transition:
one can see a slight increase of line width with energy.
 Bottom panel: same as top panel, for the difference in peak velocity
between each $K$ line and the $(6-5)$ $K=2$ line. Unlike the FHWM, no trend
is present and the
 peak velocity is approximately the same for all transitions}
\label{pend}
\end{figure}

To this purpose, we have relaxed the assumption of equal line FWHM and peak
velocity used in the line fits illustrated in Sect.~\ref{sfit} and we have
fitted each $K$ component independently from the others. 
Then we have considered the variation of the line FWHM and peak
velocity as a function of excitation energy.
To this purpose, for each source we have computed the
 ratio between the FWHM of each line and that of a reference line, and the
difference between the velocity of each line and that of the same reference
line. The latter was the (6--5) $K$=2 transition, which is strong
and does not overlap with other $K$ lines.
Such ratios and differences have been plotted as a function of the ratio
between the energies of the corresponding transitions and that of the
reference line: the
result is shown in Fig.~\ref{pend}. Clearly, no trend is
seen for the velocity:
this suggests that all \ace\
lines trace the whole emitting region, as expected for
optically thin transitions. On the contrary, the FWHM ratio shows a
slight tendency to increase with the energy of the transition.

This fact cannot be explained with
thermal broadening of the lines, because temperature changes at most from
$\sim$10~K on the clump surface to $\sim$100~K in the inner regions: the
corresponding contribution to the line width is respectively 0.1~\kms\ and
0.34~\kms, which for a non-thermal line width of at least $\sim$3~\kms\
determines a variation of
0.6\% of the line width. The latter is negligible to our purposes.

Another explanation for the observed line broadening at high energy might be
an increase of optical depth for the high-energy transitions.
It is possible to demonstrate that for an optical depth
$\tau$ at the line centre, the ratio between the line FWHM, $\Delta v$, and
the corresponding value in the optically thin limit, $\Delta v_0$ (i.e. the
intrinsic line width), is given by the expression
\begin{equation}
\frac{\Delta v}{\Delta v_0} = \sqrt{-\frac{1}{\ln 2} \,
   \ln\left[\frac{\ln 2-\ln\left(1+{\rm e}^{-\tau}\right)}{\tau}\right]}
\end{equation}
From this, one can see that $\tau\simeq4$ is needed to obtain
$\Delta v/\Delta v_0\simeq1.5$, i.e. the maximum ratio seen in the
top panel of
Fig.~\ref{pend}. As already discussed at length in Sects.~\ref{stempden}
and~\ref{stau}, such a large value must refer to a very small region,
which cannot contribute to the \ace\ emission observed by us on a much
larger scale. Therefore, line broadening
is unlikely to be due to optical depth effects.

In conclusion, we believe that the most plausible
explanation for Fig.~\ref{pend} is that the velocity
dispersion
increases with energy of the transition and hence towards the centre
of the clumps.
Various mechanisms can explain this trend:
keplerian rotation, infall motions, and turbulence
caused by winds/outflows from the embedded, young high-mass stars. Although
the present observations cannot discriminate among these, it is worth noting
that the infall hypothesis is consistent with the scenario proposed in
Sects.~\ref{sstab} and~\ref{sdens}. 

\section{Conclusions}
\label{sconc}

We have used the IRAM 30-m and FCRAO 14-m telescopes to observe a sample of
12 UC \HII\ regions in the $J$=6--5, 8--7, and 13--12 transitions of \ace.
Emission has been detected in all of the objects and the following
results have been obtained:
\begin{itemize}
\item we have estimated the diameters of the \ace\ emitting regions by comparing
 the main beam brightness temperatures measured with the 30-m and the FCRAO
 telescopes: a typical diameter turns out to be $\sim 0.9$ pc, of the same
 order of those obtained with other medium density tracers
 (Cesaroni et al. 1991; Hofner et al. 2000);
\item we have assessed that the \ace\ molecule is in LTE and derived
 an accurate
 estimate of the temperature of the molecular clumps associated with these
 high-mass star forming regions: the temperatures range from 30 to 56~K;
\item we have derived \ace\ column densities of
 $10^{14}$--$10^{15}$~cm$^{-2}$ corresponding to H$_2$ volume densities
 of $\sim$$10^5$--$10^6$~cm$^{-3}$;
\item by comparing our results with those obtained by 
 other authors using different tracers (Cesaroni et al. 1991;
 Hofner et al. 2000; Kurtz et al. 2000) we find
 that temperature and density depend on the distance from the clump
 centre, $R$, according to the relations
 $n\propto R^{-2.6}$ and $T\propto R^{-0.5}$: these are consistent with
 marginally stable clumps heated by the embedded massive stars;
\item the masses of the clumps turn out to be $\sim$3 times greater than the
 corresponding virial estimates obtained ignoring magnetic fields and external
 pressure: we suggest that this might indicate that
 the clumps are on the verge of gravitational collapse and
 that magnetic fields might play an important role in stabilising the clumps;
 we find that the magnetic fields needed for virial equilibrium
 are of the order of 1~mG;
\item the dynamical timescales on which the clumps may collapse are of
 $\sim10^5$~yr: these imply mass accretion rates of
 $\sim10^{-2}~M_\odot$\,yr$^{-1}$.
\item the line width shows a tendency to increase for increasing excitation
 energy, and hence for decreasing distance from the centre:
 a plausible explanation is that such a line broadening is due to collapse
 of the inner regions.
\end{itemize}   

\begin{acknowledgements}
It is a pleasure to thank Malcolm Walmsley for critically
reading the manuscript and Daniele Galli and Gianni Comoretto for stimulating
discussions. The anonymous referee is also acknowledged for constructive
criticisms that much improved the presentation.
\end{acknowledgements}

\end{document}